\documentclass[aps,prl,amsmath,amssymb,floatfix,twocolumn,amsmath,superscriptaddress,twocolumn,nofootinbib,tighten,letterpaper]{revtex4}
\usepackage{multirow}
\usepackage{subfigure}
\usepackage{color}
\usepackage{mathrsfs}
\usepackage{hyperref}
\usepackage[normalem]{ulem}
\usepackage{bm}

\usepackage{amssymb}   
\usepackage{amsmath}
\renewcommand\vec[1]{\ensuremath\boldsymbol{#1}} 

\usepackage{amsfonts, relsize, color}
\usepackage{graphics}
\usepackage{graphicx}
\usepackage{subfigure}
\usepackage{hyperref}
\usepackage{color}
\usepackage{comment}

\begin{document}

\title{Non-Hermitian dislocation modes: Stability and melting across exceptional points}

\author{Archisman Panigrahi}
\affiliation{Indian Institute of Science, Bangalore 560012, India}

\author{Roderich Moessner}
\affiliation{Max-Planck-Institut f\"{u}r Physik komplexer Systeme, N\"{o}thnitzer Stra{\ss}e 38, 01187 Dresden, Germany}

\author{Bitan Roy}\thanks{Corresponding author: bitan.roy@lehigh.edu}
\affiliation{Department of Physics, Lehigh University, Bethlehem, Pennsylvania, 18015, USA}

\date{\today}

\begin{abstract}
The traditional bulk-boundary correspondence assuring robust gapless modes at the edges and surfaces of insulating and nodal topological materials gets masked in non-Hermitian (NH) systems by the skin effect, manifesting an accumulation of a macroscopic number of states near such interfaces. Here we show that dislocation lattice defects are immune to such skin effect or at most display a \emph{weak} skin effect (depending on its relative orientation with the Burgers vector), and as such they support robust topological modes in the bulk of a NH system, specifically when the parent Hermitian phase features band inversion at a finite momentum. However, the dislocation modes gradually lose their support at their core when the system approaches an exceptional point, and finally melt into the boundary of the system across the NH band gap closing. We explicitly demonstrate these findings for a two-dimensional NH Chern insulator, thereby establishing that dislocation lattice defects can be instrumental to experimentally probe pristine NH topology.    
\end{abstract}

\maketitle

\begin{figure}[t!]
\includegraphics[width=0.975\linewidth]{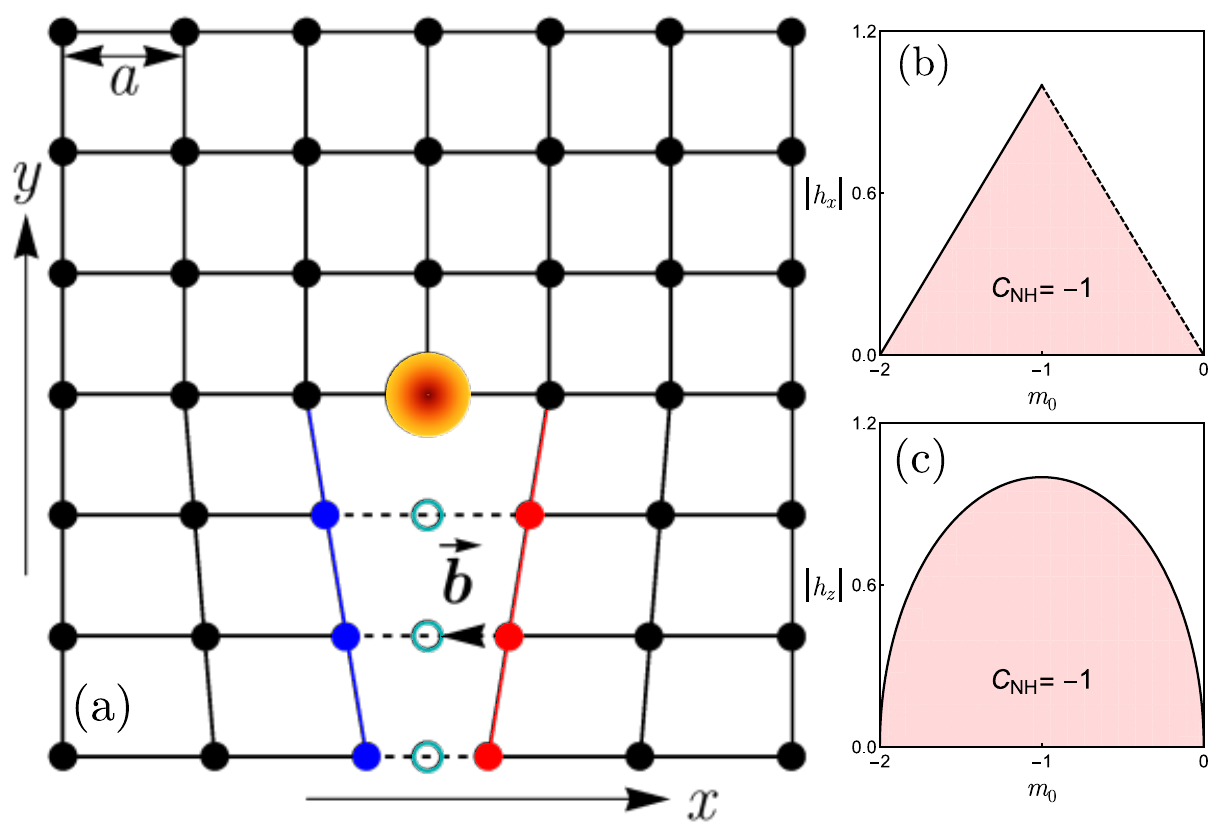}
\caption{(a) Construction of an edge dislocation with Burgers vector ${\bf b}= a {\bf e}_x$ through the Volterra cut-and-paste procedure on a square lattice by removing a line of atoms (cyan circles), ending at its center or core (orange circle) and subsequently reconnecting (dashed lines) the sites (blue and red dots) living on the edges (blue and red lines) across it. Thus, the dislocation core does not host `\emph{naked}' edges, unlike the perimeter of the system. Phase diagrams in the presence of the NH perturbation (b) $h_x$ (and $h_y$) and (c) $h_z$ for $t=t_0=1$ [Eq.~(\ref{eq:dvec})]. The eigenvalue spectra are gapped only in the red shaded region, where the NH Chern number $C_{\rm NH}=-1$ [Eq.~(\ref{eq:NHChernNumber})] and system supports topologically protected dislocation modes [Figs.~\ref{fig:PBCdislocation} and~\ref{fig:OBCdislocation}]. EPs first appear in the BZ along the black lines (defining $h^c_x$) through the ${\rm X}=(\pi,0)/a$ and ${\rm Y}=(0,\pi)/a$ points (dashed line) or the ${\rm M}=(\pi,\pi)/a$ point (solid line) in (b). In (c), EPs enter the BZ at momentum satisfying $\cos(k_x)=\cos(k_y)=m_0/2$ for $h_z=h^c_z$ (black solid line). Above the black lines EPs move in the BZ before they collide and disappear (not shown explicitly). 
}~\label{fig:dislocationsetup}
\end{figure}

\emph{Introduction}.~Breaking of translational symmetry plays a pivotal role in the experimental identification of topological materials. As such topological insulators, superconductors and semimetals host robust gapless modes at their interfaces with vacuum, encoding the bulk-boundary correspondence~\cite{Hasan-Kane-RMP, Qi-Zhang-RMP, Schnyder-RMP, Armitage-RMP}. However, this traditional approach to identify topological matter gets challenged in non-Hermitian (NH) systems, where the weight of a macroscopic number of states shifts toward a specific (but not all) boundary of the system, a phenomenon known as the \emph{skin effect}, thereby overwhelming the signature of topological boundary modes~\cite{torres:review, tanmoy:review, Bergholtz:review, Kohmoto:PRB2011, Huang:PRA2013, LFu:arXiv2017, Xu:PRL2017, Wang:PRL2018A, Wang:PRL2018B, ueda:PRX2018, ueda:PRB2018, Liangfu2018, Bergholtz:PRL2018, Berholtz:PRB2019, Ueda:NatComm2019, Lee:PRB2019, Murakami:PRL2019, Moessner:NatComm2019, LeeThomale:PRB2019, Sato:PRL2020, Vitelli:PRL2020, Xue:NatPhys2020, Fang:PRL2020, Slager:PRL2020, Song:PRB2020, hughes:PRL2021, gong:PRB2021, manna-roy:NH2022}. Therefore, identifying NH topological phases demands alternative probes, capable of circumventing the skin effect. Here we show that it can be achieved by topological lattice defects, namely dislocations [Fig.~\ref{fig:dislocationsetup}(a)], which break the translational symmetry \emph{locally} in the bulk of the system.

\begin{figure*}[t!]
\includegraphics[width=0.975\linewidth]{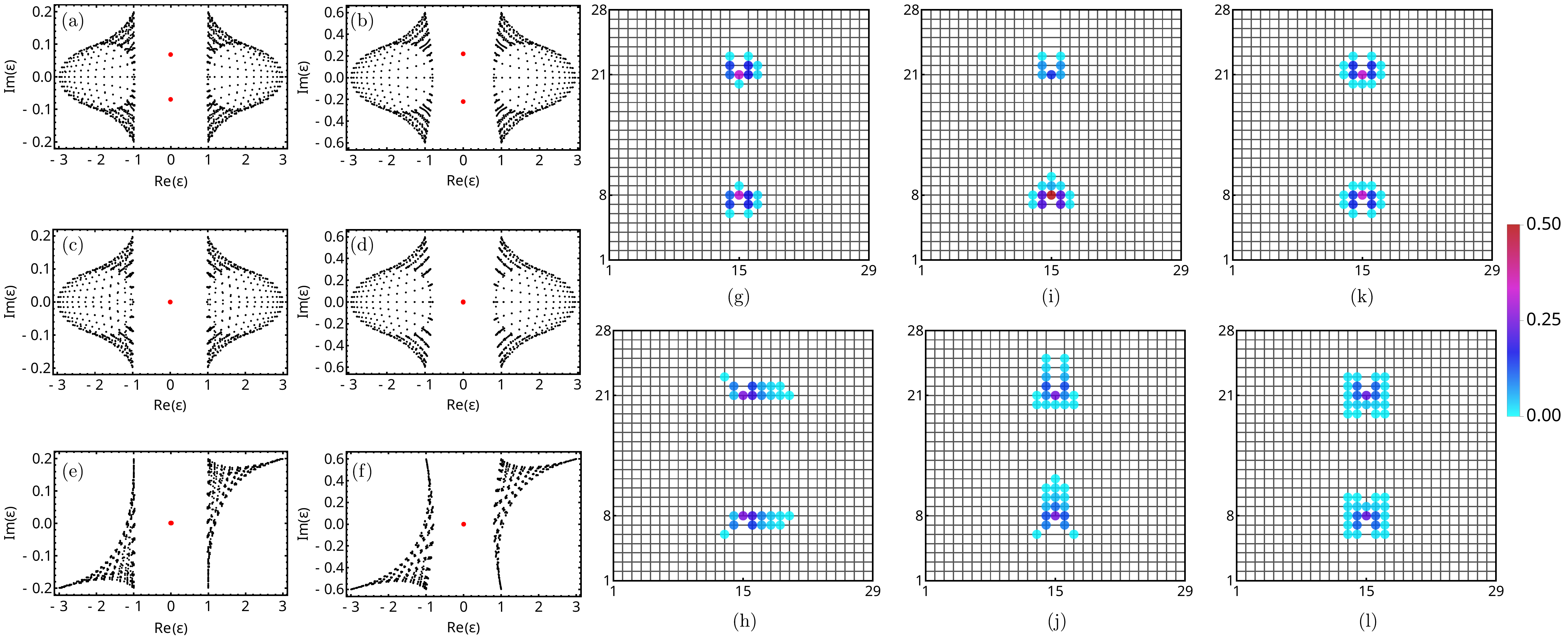}
\caption{Eigenvalues of the NH operator $H_{\rm NH}= {\boldsymbol \sigma} \cdot \vec{d}(\vec{k}) + i {\boldsymbol \sigma} \cdot \vec{h}$ in a periodic system with an edge dislocation-antidislocation pair for (a) $h_x=0.2$, (b) $h_x=0.6$, (c) $h_y=0.2$, (d) $h_y=0.6$, (e) $h_z=0.2$ and (f) $h_z=0.6$, when $t=t_0=-m_0=1$. The amplitude of the right eigenvectors corresponding to the dislocation modes [red dots in (a)-(f)] are respectively displayed in (g), (h), (i), (j), (k) and (l). Any site with its value less than $0.5 \times 10^{-4}$ is left empty. With increasing NH perturbations, the spectral gap to the rest of bulk states (black dots) decreases, and the dislocation modes gradually lose their support at their cores. For $h_{x,y,z} \geq h^c_{x,y,z}$ [white shaded region in Fig.~\ref{fig:dislocationsetup}(b) and (c)], with the appearance of EPs in the BZ, the dislocation modes melt into rest of the bulk states (as there is no skin effect in periodic systems) and disappear.               
}~\label{fig:PBCdislocation}
\end{figure*}

\begin{figure}[t!]
\includegraphics[width=1\linewidth]{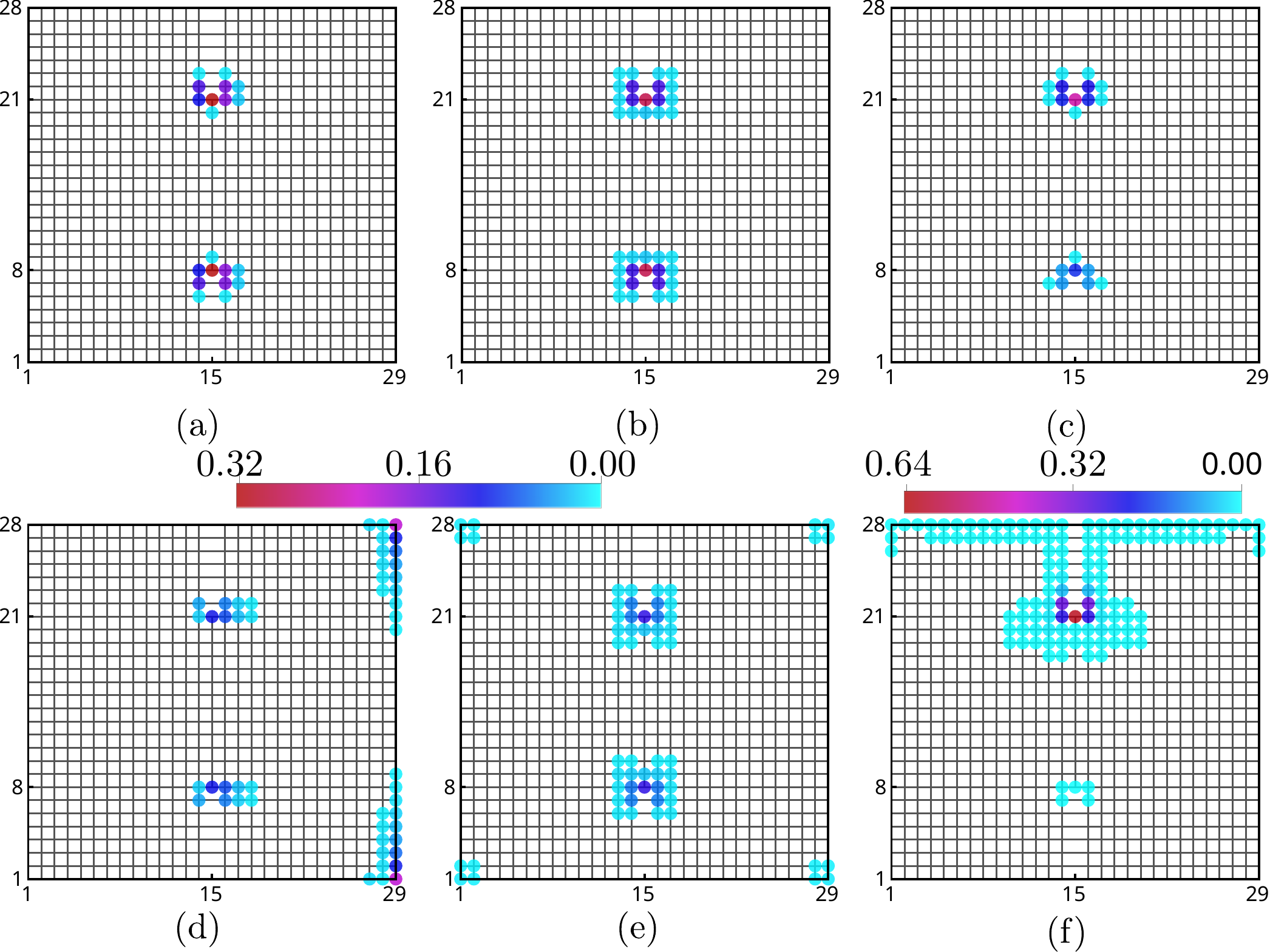}
\caption{Amplitude of right eigenvectors of the dislocation modes for (a) $h_x=0.2$, (b) $h_z=0.4$, (c) $h_y=0.05$, (d) $h_x=0.55$, (e) $h_z=0.7$ and (f) $h_y=0.45$ in systems with open boundary condition. Any site with its value less than $0.5 \times 10^{-4}$ is left empty. With increasing $h_x \; (h_y)$, dislocation modes gradually lose support at dislocation cores and their weight shifts toward the right (top) edge or `skin' of the system, due to the broken $C_4$ symmetry. Such shift with increasing $h_z$ is, however, predominantly toward the four corners, as this NH perturbation preserves the $C_4$ symmetry.
}~\label{fig:OBCdislocation}
\end{figure}

By focusing on a simple, but paradigmatic, representative of a two-dimensional (2D) Chern insulator, we show that it continues to host robust dislocation modes as NH perturbations are gradually switched on. In particular, this happens when the band inversion of the underlying Hermitian insulator occurs at the finite momentum ${\rm M}$ point in the Brillouin zone (BZ) [Figs.~\ref{fig:dislocationsetup}(b) and \ref{fig:dislocationsetup}(c)]. As such the combination of a nonzero Burgers vector of dislocations with a finite momentum of the band inversion (${\bf K}_{\rm inv}$) provides a nontrivial hopping phase, underpinning the emergence of the core state. While the dislocation modes in a NH Chern insulator reflect the nontrivial bulk topological invariant, the NH Chern number $C_{\rm NH}=-1$, their robustness stems from the (pseudo) particle-hole and crystal symmetries. Dislocation modes remain stable until the system develops exceptional points (EPs), as the modes then completely lose support at the defect core. This observation, even though most crisp in periodic systems [Fig.~\ref{fig:PBCdislocation}], nevertheless corroborates the bulk-boundary correspondence and skin effect in systems with open boundary condition. Specifically, as the NH insulator approaches a band gap closing, the weight of the dislocation modes gradually shifts toward the boundary (``skin'') of the system [Fig.~\ref{fig:OBCdislocation}]. Finally with the appearance of EPs in the BZ, when $C_{\rm NH}$ ceases to be a bonafide topological invariant, the dislocation modes completely dissolve, beyond which typically skin effect prevails.

\begin{figure}[t!]
\includegraphics[width=0.95 \linewidth]{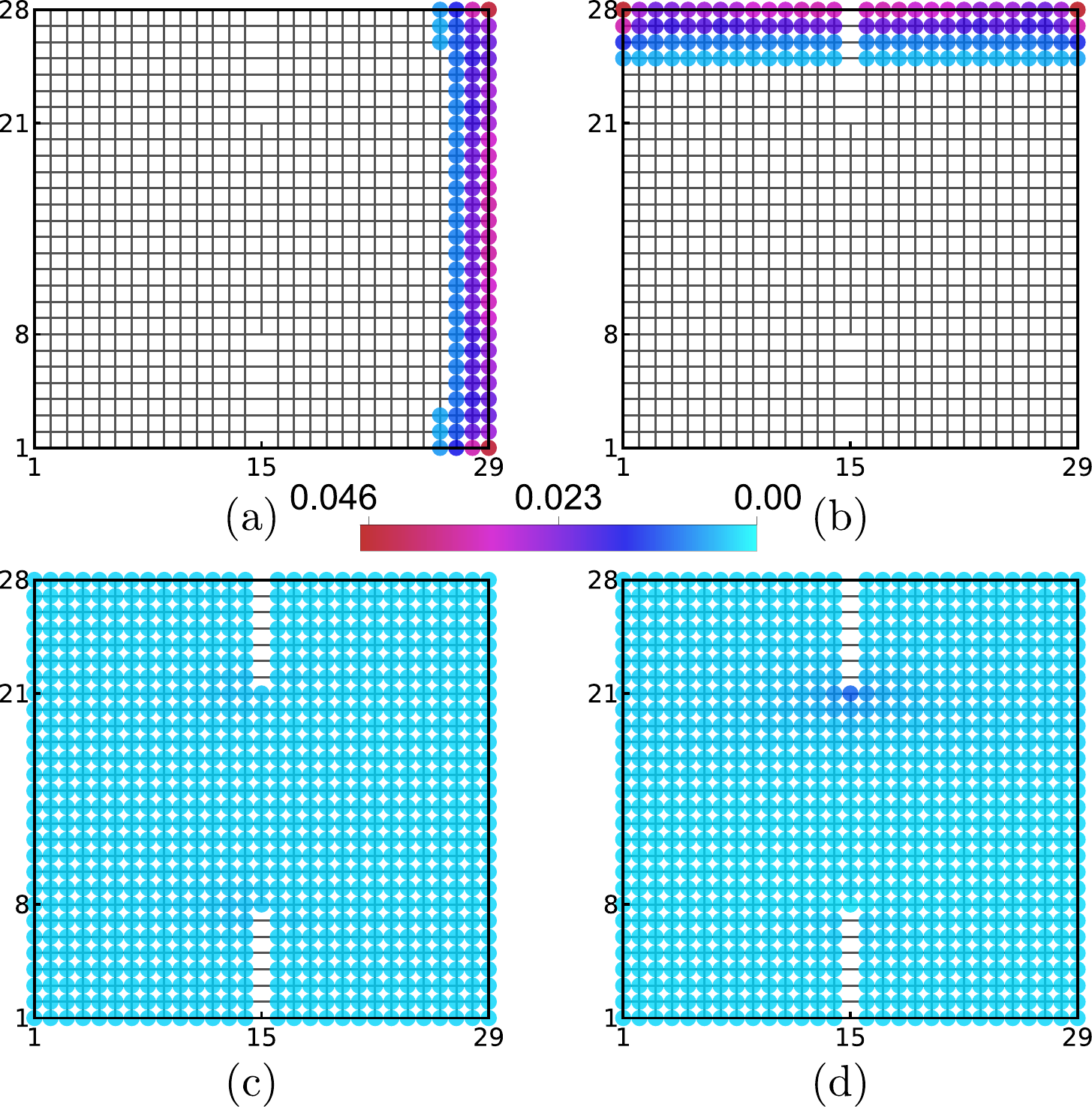}
\caption{Amplitude of all the right eigenvectors except for the dislocation modes (shown in Fig.~\ref{fig:OBCdislocation}) for $h_x=0.4$ [(a) and (c)] and $h_y=0.4$ [(b) and (d)] in systems with open [(a) and (b)] and periodic [(c) and (d)] boundary conditions. Any site with its value less than $0.3 \times 10^{-3}$ is left empty in (a) and (b). Therefore, in open boundary system the dislocation core is always devoid of skin effect. By contrast in periodic system dislocation core is does not show skin effect for $h_x$, while it features a \emph{weak} skin effect for $h_y$, whose amplitude is at least an order magnitude smaller than that in open boundary system and the dislocation modes. This is so because the associated Burgers vector ${\bf b}= a {\bf e}_x$ is parallel (orthogonal) to the direction of the skin effect for $h_x \; (h_y)$ [see text].  
}~\label{fig:dislocationSE}
\end{figure}

Finally, we show that dislocation core is always devoid of skin effect in systems with open boundaries [Figs.~\ref{fig:dislocationSE}(a) and ~\ref{fig:dislocationSE}(b)]. In periodic systems, the defect core shows no skin effect and a weak skin effect when the Burgers vector is respectively parallel and perpendicular to the direction of the skin effect in open boundary systems, as shown in Fig.~\ref{fig:dislocationSE}(c) and \ref{fig:dislocationSE}(d). However, the magnitude of the \emph{weak} skin effect is at least an order magnitude weaker than that for dislocation bound modes. Therefore, dislocation defects are instrumental for experimentally probing pristine NH topology in real materials.

Recently, NH topological phases have been realized in photonic crystals with gain and/or loss~\cite{NH-photonic:1, NH-photonic:2, NH-photonic:3, NH-photonic:4, NH-photonic:5, NH-photonic:6} and in mechanical metamaterials~\cite{NH-mechanical:1, NH-mechanical:2, NH-mechanical:3, NH-mechanical:4}. On these platforms lattice defects have been engineered to probe Hermitian topological phases~\cite{defect-photonic:1, defect-photonic:2, defect-photonic:3, defect-mechanical:1, defect-mechanical:2, defect-mechanical:3}. Therefore, our predictions can, in principle, be tested in NH metamaterials. Consult ``Summary and discussions'' for detailed discussion.

\emph{Model}.~In the absence of any NH perturbation, the system we study is described by the Hamiltonian $H= {\boldsymbol \sigma} \cdot \vec{d}(\vec{k})$, where ${\boldsymbol \sigma}$ is the vector of Pauli matrices and 
\begin{align}~\label{eq:dvec}
\hspace{-0.54cm}\vec{d}(\vec{k})= \bigg( t \sin (k_x a), t \sin (k_y a), t_0 \sum_{j=x,y} \cos(k_j a) -m_0 \bigg),
\end{align}
with momentum $\vec{k} \in [-\pi,\pi]/a$. This form of the $\vec{d}$-vector describes Chern and normal insulators for $|m_0/t_0|<2$ and $|m_0/t_0|>2$, respectively. In the topological regime, there exist two distinct phases for $0<m_0/t_0<2$ and $-2<m_0/t_0<0$, with band inversions respectively at the $\Gamma=(0,0)$ and ${\rm M}=(\pi,\pi)/a$ points of the BZ. These two phases are characterized by distinct values of the first Chern number ($C$)~\cite{TKNN}, with $C=1$ in the $\Gamma$ phase and $C=-1$ in the ${\rm M}$ phase. Both phases support a topological chiral edge mode, reflecting the nontrivial value of the bulk topological invariant $C$. However, as we will see, only the latter supports zero energy modes when a dislocation is created in the bulk of the system, due to the band inversion occurring at finite momentum ${\rm M}$~\cite{juricic-PRL}.

\emph{Dislocations}.~A dislocation on a 2D lattice is created by removing a line of atoms up to a site, known as its core or center, and subsequently joining the sites across the line of missing atoms. As a result, translational symmetry gets restored everywhere away from the close vicinity of its core. Any closed path around the dislocation core therefore features a missing translation by the Burgers vector ${\bf b}$, with ${\bf b}=a {\bf e}_x$ in Fig.~\ref{fig:dislocationsetup}. Consequently, an electron with momentum ${\bf K}$ encircling the dislocation core picks up a hopping phase $\exp[i \Phi]$, where $\Phi={\bf K} \cdot {\bf b}$ (modulo $2\pi$). In topological materials with the band inversion at momentum ${\bf K}_{\rm inv}$, this phase is given by $\Phi={\bf K}_{\rm inv} \cdot {\bf b}$~\cite{ran-zhang-vishwanath, teo-kane, juricic-PRL, nagaosa, juricic-natphys, hughes-yao-qi, you-cho-hughes, roy-juricic-dislocation, nag-roy:floquetdislocation}. Hence, in the ${\rm M}$ phase $\Phi=\pi$ (nontrivial), while $\Phi=0$ (trivial) in the $\Gamma$ phase.

After reconnecting the sites on the edges located across the line of missing atoms, the corresponding edge modes hybridize and their energies repel. Such a level repulsion is represented by a trivial or uniform Dirac mass in the $\Gamma$ phase, as $\Phi=0$ therein. Consequently, the edge modes are trivially gapped and the dislocation core does not support any topological mode in the $\Gamma$ phase. On the other hand, hybridization between the one-dimensional edge modes is captured by a \emph{domain wall} mass in the ${\rm M}$ phase, as $\Phi=\pi$ therein. As a result, the dislocation core in Hermitian systems hosts a zero energy topological midgap state at its core in the ${\rm M}$ phase, following the Jackiw-Rebbi mechanism~\cite{jackiw-rebbi}. For the rest of the discussion we, therefore, focus on the ${\rm M}$ phase, and address the stability and dissolution of dislocation modes in the presence of NH perturbations.

\emph{Symmetries}.~In Hermitian systems, the dislocation mode gets pinned at zero energy due to an antiunitary particle-hole symmetry of the Hamiltonian $\Theta H \Theta^{-1}=-H$, where $\Theta=\sigma_x {\mathcal K}$ and ${\mathcal K}$ is the complex conjugation. Therefore, any pair of eigenmodes at energies $\pm E$, denoted by $| \pm E \rangle$, are related to each other by $\Theta | \pm E \rangle = |\mp E \rangle$. Hence, zero energy modes can be chosen to be eigenstates of the operator $\Theta$~\cite{broy-antiunitary}. Although not directly pertinent for the stability of the Chern insulators, the above model enjoys a four-fold ($C_4$) rotational symmetry, generated by ${\mathcal R}_4=\exp[i \pi \sigma_3/4]$, under which $\vec{k} \to (-k_y,k_x)$, about which more in a moment.

\emph{NH system}.~A generic NH perturbation in this system is captured by the antiHermitian operator $H_{\rm NH}= i {\boldsymbol \sigma} \cdot \vec{h}$, where $i =\sqrt{-1}$ and the vector $\vec{h}=(h_x,h_y,h_z)$ is real. Next we analyze the dislocation modes of the NH operator $H+ H_{\rm NH}$, starting from the ${\rm M}$ phase of the Hermitian Chern insulator. In what follows we scrutinize the role of $h_x$, $h_y$ and $h_z$ separately. The eigenvalues ($\varepsilon$) of any NH operator are generically complex ($\varepsilon \neq \varepsilon^\star$), and the associated left and right eigenvectors are generically not Hermitian conjugates of each other, i.e., $_{_L}\langle \varepsilon | \neq | \varepsilon \rangle_{_R}^\dagger$.

\emph{Phase diagram  and topological invariant}.~The NH operator $H^x_{\rm NH}=H + i \sigma_x h_x$ preserves the particle-hole symmetry $\Theta H^x_{\rm NH} \Theta^{-1}=-H^x_{\rm NH}$. But it breaks the $C_4$ symmetry. The phase diagram of $H^x_{\rm NH}$ is shown in Fig.~\ref{fig:dislocationsetup}(b). The red shaded region is occupied by a NH Chern insulator, characterized by $C_{\rm NH}=-1$, defined as~\cite{supplementary} 
\begin{eqnarray}~\label{eq:NHChernNumber}
C_{\rm NH}= {\rm Re} \int \frac{d^2 \vec{k}}{4\pi} \left[ \partial_{x} \hat{\vec{d}}_{\rm NH}(\vec{k}) \times \partial_{y} \hat{\vec{d}}_{\rm NH}(\vec{k}) \right] \cdot \hat{\vec{d}}_{\rm NH}(\vec{k}),
\end{eqnarray}
where $\partial_j \equiv \partial_{k_j}$, $\vec{d}_{\rm NH}(\vec{k})=\vec{d}(\vec{k})+ i \vec{h}$, and $\hat{\vec{d}}_{\rm NH}(\vec{k})=\vec{d}_{\rm NH}(\vec{k})/\sqrt{\vec{d}^2_{\rm NH}(\vec{k})}$. For $\vec{h}=0$, we recover the standard definition of the Chern number~\cite{TKNN}. 

The EPs enter the BZ at the black lines and persist in the white shaded regime, where $C_{\rm NH}$ is no more a bonafide bulk topological invariant~\cite{ueda:PRB2018, Liangfu2018}. For $t=t_0=1$, the phase boundary between the insulating and nodal NH phases is determined by $|h_x|=h^c_x=|m_0|$ for $-1 \leq m_0 \leq 0$, along which the EPs enter the BZ at the ${\rm X}$ and ${\rm Y}$ points. By contrast, $h^c_x=|m_0+2|$ when $-2 \leq m_0 \leq -1$ where the EPs enter at the ${\rm M}$ point~\cite{supplementary}. Above these lines the EPs move in the BZ. Finally they collide and disappear~\cite{ueda:PRB2018}. The phase diagram of the NH operator $H^y_{\rm NH}=H+i \sigma_y h_y$ that also preserves the particle-hole symmetry $\Theta H^y_{\rm NH} \Theta^{-1}=-H^y_{\rm NH}$ and breaks the $C_4$ symmetry, is identical to Fig.~\ref{fig:dislocationsetup}(b).

The NH operator $H^z_{\rm NH}=H+i \sigma_z h_z$, on the other hand, preserves the $C_4$ symmetry and satisfies $\Theta H^z_{\rm NH} \Theta^{-1}=-(H^z_{\rm NH})^\dagger$, which we name \emph{pseudo} particle-hole symmetry~\cite{pseudoPH:Comment}. A NH Chern insulator with $C_{\rm NH}=-1$ then occupies the red shaded region of the phase diagram in Fig.~\ref{fig:dislocationsetup}(c). EPs enter the BZ at the black solid line, determined by $|h_z|=h^c_z=\left[ 2|m_0|-m^2_0 \right]^{1/2}$ for $-2 \leq m_0 \leq 0$ and $t=t_0=1$. They appear at points connected by four-fold rotations and satisfying $\cos(k_x)=\cos(k_y)=m_0/2$.

Phase diagram of NH models can alternatively be derived from non-Bloch band theory~\cite{Wang:PRL2018A, Wang:PRL2018B}, which although typically is different from the one computed from NH Bloch band theory, can crucially depend on the geometry of the system in dimensions $d \geq 2$~\cite{chenfang:natcomm2022}. Nonetheless, the phase diagrams in Figs.~\ref{fig:dislocationsetup}(b) and (c) in terms of the NH Chern number ($C_{\rm NH}$) predicts the correct behavior of the dislocation modes, which we discuss next, in the sense that these modes are found in the entire red shaded regimes with periodic and open boundary conditions, only where $C_{\rm NH}$ is quantized to $-1$ and the band inversion takes place around the ${\rm M}$ point of the BZ~\cite{comment:chernnumber}. See Fig.~S9~\cite{supplementary}. We note that for stronger NH coupling, where $C_{\rm NH}$ cease to be a bonafide topological invariant as the energy spectra display point gap, with no analog in the Hermitian systems, an ``intrinsic" topology may emerge, exclusively realized in NH systems~\cite{Sato:PRL2020}.

\emph{NH dislocation modes}.~In the presence of dislocations, the NH Chern insulators, occupying the red shaded regions in Figs.~\ref{fig:dislocationsetup}(b) and \ref{fig:dislocationsetup}(c), support topologically robust localized modes around their cores, which here we establish from the spatial profile of the amplitude of the corresponding right eigenvectors. The results are displayed in Fig.~\ref{fig:PBCdislocation} in a periodic system with an edge dislocation-antidislocation pair. However, with increasing NH perturbations, the dislocation modes spread away from its core and its weight gets gradually absorbed by other bulk states as there is no skin effect in periodic systems. These outcomes are qualitatively unchanged for the left eigenvectors, except for $h_x$ ($h_y$) their weight shifts toward the left (bottom) side of the system. Finally, with the appearance of EPs in the BZ, the dislocation modes completely melt and disappear. This observation can be qualitatively justified in the following way. The minimal value of the real component of the gap among the eigenvalues of a NH operator with ${\rm Re}(\varepsilon)>0$ and ${\rm Re} (\varepsilon)<0$ (black dots), namely $\delta{\rm Re}(\varepsilon)$, sets the localization length ($\ell_{\rm dis}$) of the dislocation modes, $\ell_{\rm dis} \sim 1/\delta{\rm Re}(\varepsilon)$. As $\delta{\rm Re}(\varepsilon)$ decreases with increasing $h_x$, $h_y$ and $h_z$ [Fig.~\ref{fig:PBCdislocation}], the dislocation modes gradually delocalize. Finally, in the presence of EPs, $\delta{\rm Re}(\varepsilon)=0$: dislocations can no longer support localized modes as $\ell_{\rm dis} \to \infty$.

Melting of the dislocation modes takes place through the skin of the systems with open boundary condition. For sufficiently weak NH perturbations these modes are strongly localized in the interior of the system, as shown in Figs.~\ref{fig:OBCdislocation}(a), \ref{fig:OBCdislocation}(b) and \ref{fig:OBCdislocation}(c) for $h_x \ll h^c_x$, $h_z \ll h^c_z$ and $h_y \ll h^c_y$, respectively. As $\vec{h} \to \vec{h}^c$, the weight of the right eigenvectors for the dislocation modes gradually shifts toward the `skin' of the system [Figs.~\ref{fig:OBCdislocation}(d)-(f)]. However, due to the finite size effects they disappear from the system even when $\vec{h} \lesssim \vec{h}^c$. The weight of the right eigenvectors only accumulates along the right/top edge of the system, manifesting the broken $C_4$ symmetry by $H^{x/y}_{\rm NH}$. On the other hand, the left eigenvectors populate the left/bottom edge of the system and the biorthogonal product of the dislocation modes does not appear near any skin~\cite{supplementary}. By contrast, four corners, connected via the $C_4$ rotations, are \emph{equally} populated in the presence of $h_z$. This feature remains qualitatively unchanged for the left eigenvectors and the biorthogonal product~\cite{supplementary}.

 Finally, we address the particle-hole symmetry of the dislocation modes. For the NH perturbation $h_x$ or $h_y$, the dislocation modes appear at purely imaginary eigenvalues with $\varepsilon=-\varepsilon^\star$. Consequently $\Theta|\varepsilon \rangle_{_R}=|\varepsilon \rangle_{_R}$ up to an overall unimportant phase, for any purely imaginary $\varepsilon$, and the right eigenvectors are eigenstates of the particle-hole operator $\Theta$. Similar conclusions hold for the left eigenvectors~\cite{supplementary}. For the NH perturbation $h_z$ the dislocation modes possess complex eigenvalues, for which $\Theta |\varepsilon \rangle_{_R}= |-\varepsilon \rangle_{_L}$, due to the pseudo particle-hole symmetry. As the system does not permit any additional $C_4$ symmetry preserving NH perturbation, the left and right dislocation modes, despite not being eigenstates of $\Theta$, cannot be mixed with other states. Therefore, irrespective of the boundary conditions stable and topologically robust dislocation modes are found in a NH Chern insulator with $C_{\rm NH}=-1$. Dislocation lattice defects this way capture the bulk-boundary correspondence.

\emph{Dislocation skin effect}.~In a system with an open boundary condition there is no skin effect at the dislocation core [Figs.~\ref{fig:dislocationSE}(a) and \ref{fig:dislocationSE}(b)]. The situation is subtle in periodic systems. Notice that the dislocation core [orange site in Fig.~\ref{fig:dislocationsetup}(a)] has 2 (1) nearest-neighbor (NN) sites in the $x \; (y)$ direction when the Burgers vector is ${\bf b}= a {\bf e}_x$, like the sites constituting the top and bottom edges in the $y$ direction, accommodating NH skin effect for $h_y$. By contrast, $h_x$ displays NH skin effect along $x$ directional left and right edges, constituted by sites with 1 (2) NN sites in the $x \; (y)$ direction. Consequently, when we impose periodic boundary condition to eliminate outer naked edges, there is no skin effect for $h_x$ at dislocation core. But with $h_y$ a \emph{weak} skin effect is observed therein, whose amplitude is, however, at least an order magnitude smaller than that at the outer naked edges and for dislocation modes. See Figs.~\ref{fig:dislocationSE}(c) and \ref{fig:dislocationSE}(d). The absence/presence of dislocation skin effect with $h_x/h_y$ gets reversed when ${\bf b}= a {\bf e}_y$. With $h_z$ there is no skin effect anywhere in the system irrespective of the boundary condition and ${\bf b}$. Hence, dislocation lattice defects stands as a probe for pristine NH topology in real materials.

\emph{Summary and discussions}.~To summarize, here we show that the interplay between the Burgers vector of dislocation lattice defect and the finite ${\bf K}_{\rm inv}$ gives rise to stable localized modes at their cores in a NH Chern insulator. However, with increasing NH perturbations these modes slowly lose support at the dislocation core and with the appearance of EPs they melt into the ``skin'' (other bulk states) of a system with an open (periodic) boundary condition. Existence of these modes are guaranteed by a nontrivial integer bulk topological invariant, the NH Chern number ($C_{\rm NH}$), while they are forbidden to mix with other states due to the (pseudo) particle-hole and crystalline $C_4$ symmetries. Thus, NH dislocation modes by virtue of being buried deep inside the bulk bypass the skin effect in open boundary systems or occasionally at most encounters a \emph{weak} skin effect in periodic system and therefore unveil \emph{pristine} signatures of NH topology. In the future, this construction can be extended to 3D NH topological systems in the presence of edge and screw dislocations, possibly hosting 1D NH helical metal in their cores.

Metamaterials, such as photonic~\cite{review:photonic1} and phononic or mechanical~\cite{review:mechanical1,review:mechanical2} lattices, constitute the most promising platform where our predictions can be tested. In photonic crystals lattice defects have already been engineered~\cite{defect-photonic:1, defect-photonic:2}. In the presence of gain and/or loss, they yield a NH setup, where topological edge states, bulk EPs and Fermi arcs connecting them have been observed~\cite{NH-photonic:1, NH-photonic:2, NH-photonic:3, NH-photonic:4, NH-photonic:5, NH-photonic:6}. In photonic lattices dislocation defects can be realized through the requisite $\pi$ hopping phase modification across the line of missing waveguides~\cite{hafezi}, and dislocation modes can be detected via a two point pump probe~\cite{defect-photonic:2}, reflection spectroscopy~\cite{defect-photonic:3}, and real-space mapping of isofrequency contours from angle-resolved scattering measurements~\cite{NH-photonic:3}. Signatures of NH topology, such as edge states and EPs, have been observed in mechanical systems as well~\cite{NH-mechanical:1, NH-mechanical:2, NH-mechanical:3, NH-mechanical:4}, where at least the right eigenstates can be probed directly~\cite{NH-mechanical:3, NH-mechanical:4}. In mechanical systems both dislocation and disclination lattice defects have been engineered to probe Hermitian topological phases~\cite{defect-mechanical:1, defect-mechanical:2, defect-mechanical:3}, by introducing hopping phase modulation of $\pi$ across the line of missing microwave resonators (playing the role of lattice sites). Besides directly probing right eigenvectors, dislocation modes can also be detected from the zeros of the reflection coefficient~\cite{NH-mechanical:2} or mechanical susceptibility~\cite{defect-mechanical:1}. Since NH dislocation modes can be detected from the right or left eigenvectors or their biorthogonal product~\cite{supplementary}, we expect that in the future lattice defects can be instrumental in probing the underlying NH topology, at least in topological metamaterials.

\emph{Note added}. After posting the paper on arXiv we became aware of two works on dislocation NH skin effects~\cite{dislocationNH:1, dislocationNH:2}. However, none of them discussed dislocation bound modes nor addressed the direction dependence of \emph{weak} skin effect at defect core in periodic systems.

\emph{Acknowledgments}.~A.P. acknowledges support from the KVPY programme. B.R. was supported by a startup grant from Lehigh University. We acknowledge support from the Deutsche Forschungsgemeinschaft through SFB 1143 (Project-id No.~247310070) and cluster of excellence ct.qmat (EXC 2147, Project-id No.~390858490). We thank Emil Bergholtz for useful discussions and Vladimir Juri\v ci\' c for critical reading of the manuscript. We thank two anonymous referees for their thoughtful comments.

\end{document}